# An end-to-end network slicing framework for 5G wireless communication systems

Qian (Clara) Li, Geng Wu, Apostolos (Tolis) Papathanassiou, Udayan Mukherjee
Intel Corporation, USA
Emails: {clara.q.li, geng.wu, apostolos.papathanassiou, udayan.mukherjee}@intel.com

**Abstract**

Wireless industry nowadays is facing two major challenges: 1) how to support the vertical industry applications so that to expand the wireless industry market and 2) how to further enhance device capability and user experience. In this paper, we propose a technology framework to address these challenges. The proposed technology framework is based on end-to-end vertical and horizontal slicing, where vertical slicing enables vertical industry and services and horizontal slicing improves system capacity and user experience. The technology development on vertical slicing has already started in late 4G and early 5G and is mostly focused on slicing the core network. We envision this trend to continue with the development of vertical slicing in the radio access network and the air interface. Moving beyond vertical slicing, we propose to horizontally slice the computation and communication resources to form virtual computation platforms for solving the network capacity scaling problem and enhancing device capability and user experience. In this paper, we explain the concept of vertical and horizontal slicing and illustrate the slicing techniques in the air interface, the radio access network, the core network and the computation platform. This paper aims to initiate the discussion on the long-range technology roadmap and spur development on the solutions for E2E network slicing in 5G and beyond.

1. Introduction

The wireless industry finds itself at a turning point today. On wireless communication technologies, further improving the spectral efficiency at the radio link level is becoming increasingly challenging. New ways need to be found to build future networks and devices so to meet the ever increasing capacity demand. On services and applications, as smartphone penetration is generally accelerating globally and is starting to saturate in many developed markets, new services and applications need to be exploited to connect not only human beings but more things across the industries and to make our living environment more intelligent. On user experience, a technology foundation needs to be created that enables future generations of devices and services for the coming decades that are likely beyond what we could imagine today. To achieve these goals, 5G and future generations of networks and devices are about the combination of computing and communications and about end-to-end solutions. This is a paradigm shift from previous generations where technology development focused primarily on communications.

In late 4G and early 5G, the wireless industry started to vertically slice big mobile broadband network into multiple virtual networks to serve vertical industries and applications in a more cost efficient manner. Each network slice can have different network architecture, and different application, control, packet and signal processing capacity to achieve optimum return on investment. A new industry or type of service can be added to an existing network instead of deploying a new network. Vertical network slicing practically segregates the traffic from a vertical application standpoint from the rest of general mobile broadband services, practically avoiding or dramatically simplifying a traditional QoS engineering problem (i.e., QoS mechanism can be imposed to each slice instead of imposed to all the traffics among all the slices). Vertical network slicing was primarily focused on core network nodes enabled by techniques such as network functions virtualization (NFV) and software defined networking (SDN) [1]-[3]. With time we see this trend continues and expends into the radio access networks and the air interface. Examples of vertical slices developed or under developing in 4G LTE include the machine-type communication (MTC) slice, and the narrow-band internet-of-things (NB-IOT) slice [4][5]. Adding new slices in LTE is to patch on to the baseline LTE framework which was primarily designed for mobile broadband communication. In 5G, a forward-compatible design is desirable to provision for adding new slices in the future [6].

Moving forward, in the later phase of 5G and beyond, we expect the network capacity and user experience need to be further improved. However, the capacity increase does not need to be end-to-end uniform: the capacity scaling factor can be higher when closer to a user, and lower as we move deeper into the infrastructure network as showed in Figure 1, where we depicted a case with 10,000 capacity times scaling at the very edge of the network and a 10 times scaling at the network core. Such non-uniform capacity scaling is driven by the new types of user traffics and services and is founded by communication technology fundamentals: We expect a large amount of user traffic generated at the network edge due to the ever increasing number of devices and the significantly increased device sensing capability. As device sensing is local, it is desirable to keep sensed data processing and the corresponding decisions and actions to be local so that to reduce latency and improve privacy and security. As a result, the amount of data going into the deeper network will be less and the traffic scaling will be non-uniform. The non-uniform traffic scaling requires non-uniform network capacity scaling. The feasibility of non-uniform capacity scaling has been well proved in communication theory, where studies showed as cell goes dense higher frequency and spatial reuse can be achieved and consequently higher network capacity [7].

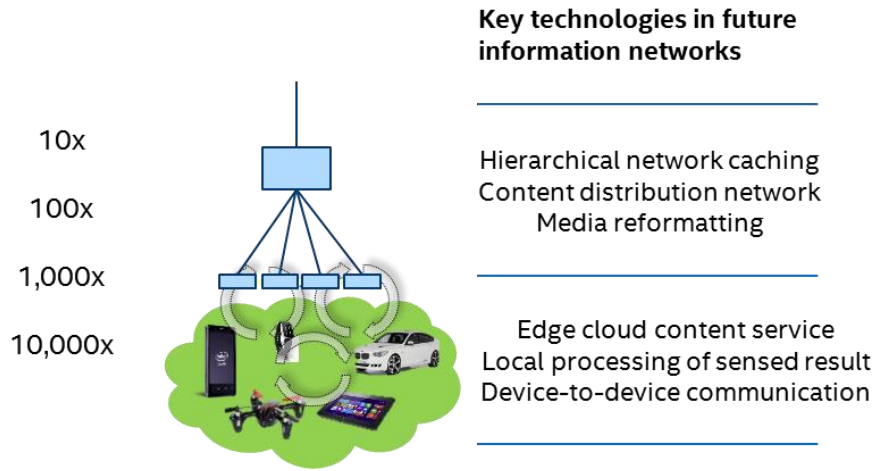

Figure 1 Illustration on non-uniform capacity scaling

The non-uniform capacity scaling can be viewed as using computation to help communication, i.e., edge computing and processing reduce traffic towards the deeper infrastructure network. The augment of device capability and the user experience enhancement, on the other hand, largely relies on using communication to help computation. Despite that the devices further shrink in size from portable devices to wearable devices, the user expectation on computation keeps increasing. We expect computation offloading will be needed to help deliver user experience, e.g., base stations slice part of their computation resource to help computation at the portable devices, portable device slice out a part of their computation resource to help the computation of the wearable devices. Computation offload helps devices to go beyond its physical limitation. The communication link connecting the donor and client ends enables computation offloading between the two.

Horizontal slicing is designed to accommodate the new trend of capacity scaling and enable edge cloud computing and computing offloading: The computing resources in the base station and the portable device (or high-capable device) will be horizontally sliced, and these slices, together with the computation resource slice of the wearable devices (or low-capable device) will be integrated to form a virtual computing platform through a new 5G air interface design to significantly augment the computing capability of future portable and wearable devices as well as perform local traffic processing. With horizontal slicing, the definition of end-to-end needs to be revisited, as the traffic flow terminates within the horizontal slice built among devices with direction communication link (and likely in close proximity).

In this paper, we first introduce the concept of network slicing and describe the system architecture with E2E vertical and horizontal network slicing, and then discuss technologies in the air interface, the RAN and the CN to enable 5G system with E2E network slicing. Then, the technology of applying horizontal slicing for enhancing device capability and user experience will be elaborated. This paper aims to initiate the discussion on the long-range technology roadmap and spur development on the solutions for E2E network slicing in 5G and beyond.

## 2. Network slicing concept and system architecture

Slicing in general is to use virtualization technology to architect, partition and organize computing and communication resources of a physical infrastructure to enable flexible support of diverse use case realizations. With network slicing, one physical network is sliced into multiple virtual networks, each architected and optimized for a specific requirement and/or specific application/service. A network slice is self-contained in terms of operation and traffic flow and can have its own network architecture, engineering mechanisms and network provision. Network slicing techniques in 5G is expected to simplify the creation and operation of network slices and allows function reuse and resource sharing of the physical network infrastructure.

As discussed in Section 1, we apply network slicing in two dimensions: Vertical network slicing and horizontal network slicing. Vertical slicing targets at supporting vertical industry and markets. It enables resource sharing among services and applications and avoids or simplifies a traditional QoS engineering problem. Horizontal slicing, as a step forward, targets at extending the capabilities of mobile devices and enhancing user experiences. Horizontal slicing goes across and beyond platforms physical boundaries. It enables resource sharing among network nodes and devices, i.e., high capable network nodes/devices share their resources (e.g., computation, communication, storage) to enhance the capabilities of less capable network nodes/devices. The end result of horizontal slicing is to spin off a new generation of underlay moving network clusters, where terminals becoming moving networking nodes. Horizontal slicing requires over-the-air resource sharing across network nodes. The 5G air interface would be an integrated part and an enabler of horizontal slicing. Horizontal slicing, in turn, differentiates and enriches 5G air interface.

Vertical slicing and horizontal slicing form independent slices. The end-to-end traffic flow in a vertical slice usually transits between the core network and the terminal devices. The end-to-end traffic flow in a horizontal slice is usually local and transits between two ends of the horizontal slice, for instance, in between a wearable device and a portable device or in between a wearable device and a small cell access point.

In vertical slicing, each of the network nodes usually implements similar functions among slices. The dynamic implement of a network node is mostly to dynamically allocation resources to each of the slices. In horizontal slicing, however, new functions could be created at a network node when supporting a slice. For example, a portable device may need different functions to support different types of wearable devices. The dynamic part could lie in the network functions as well as the resource partition.

Figure 2 illustrates the concept of vertical and horizontal network slicing. In the vertical domain, the physical computation/storage/radio processing resources in the network infrastructure (as denoted by the servers and base stations) and the physical radio resources (in terms of time, frequency, and space) are sliced to form end-to-end vertical slices via properly designed slice pairing functions. The criterion can be different when slicing the radio, the RAN and the CN. Slice pairing functions are defined to pair the radio, RAN and CN slices to form end-to-end slices for different services and applications. Figure 2shows examples of end-to-end vertical slices. The mapping between RAN slices and CN slices are not necessarily

1:1. In the horizontal domain, the physical resources (in terms of computation, storage, radio) in adjacent layers of the network hierarchy are sliced to form horizontal slices. A device could operate on multiple slices. For instance, a smart phone can operate in a vertical slice on mobile broad band (MBB) service, a vertical slice on health care service and a horizontal slice supporting wearable devices.

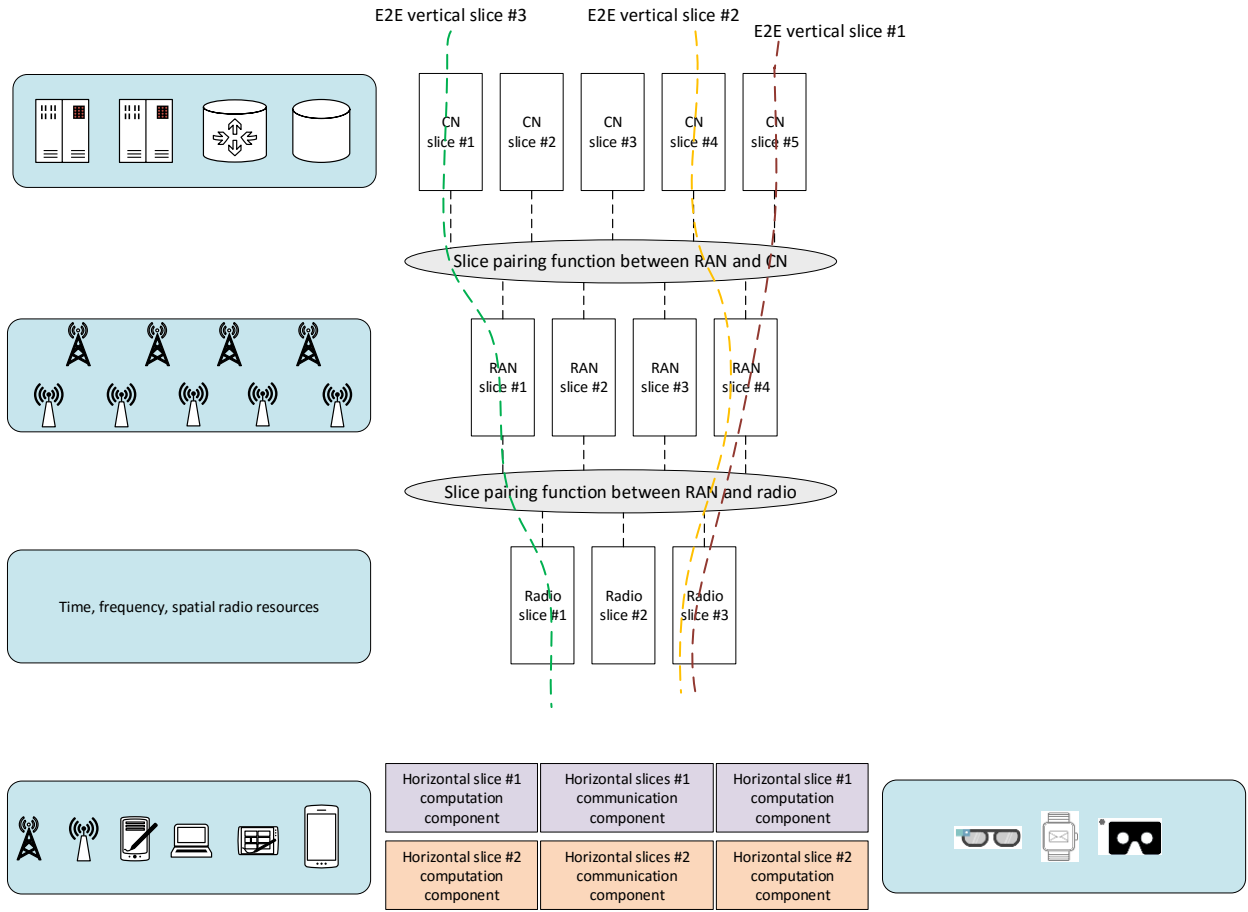

Figure 2 illustration of vertical and horizontal network slicing

## 3. Enabling vertical slicing in the air interface

Besides meeting the 5G requirements (e.g., data rate, latency, number of connections, etc.), the desirable features of the air interface to enable network slicing and in general 5G include:

- Flexibility: Support flexible radio resource allocation among slices
- Scalability: Easily scale up with the addition of new slices
- Efficiency: Efficiently use the radio and energy resources

One way to achieve the desired features is setting out from the physical (PHY) and medium access (MAC) layer architecture, as illustrated in Figure 3 (a). The basic components are a flexible partition of the PHY resources, an abstraction of the physical PHY resources into PHY resource subsets each for one of the slices, and a build of the operations of MAC and higher layers based on the PHY resource subsets. To be more specific, the physical radio resource is partitioned into multiple segments each is defined by one numerology to meet certain communication requirement, such as low latency, wide coverage, etc, where numerology refers to the values of the basic physical transmission parameters defining the radio link such as the waveform, the sampling rate, the symbol duration, the frame/subframe length, etc. Each slice occupies a subset of physical resources taken from one or multiple numerology segments. On top of the physical resource subsets, MAC operation can then be partitioned into two layers: Level-1 MAC performs intra-slice traffic multiplexing and scheduling; Level-2 MAC performs inter-slice resource allocation. The two-layer MAC partition avoids the complexity of jointly scheduling multiple slices and allows better scalability.

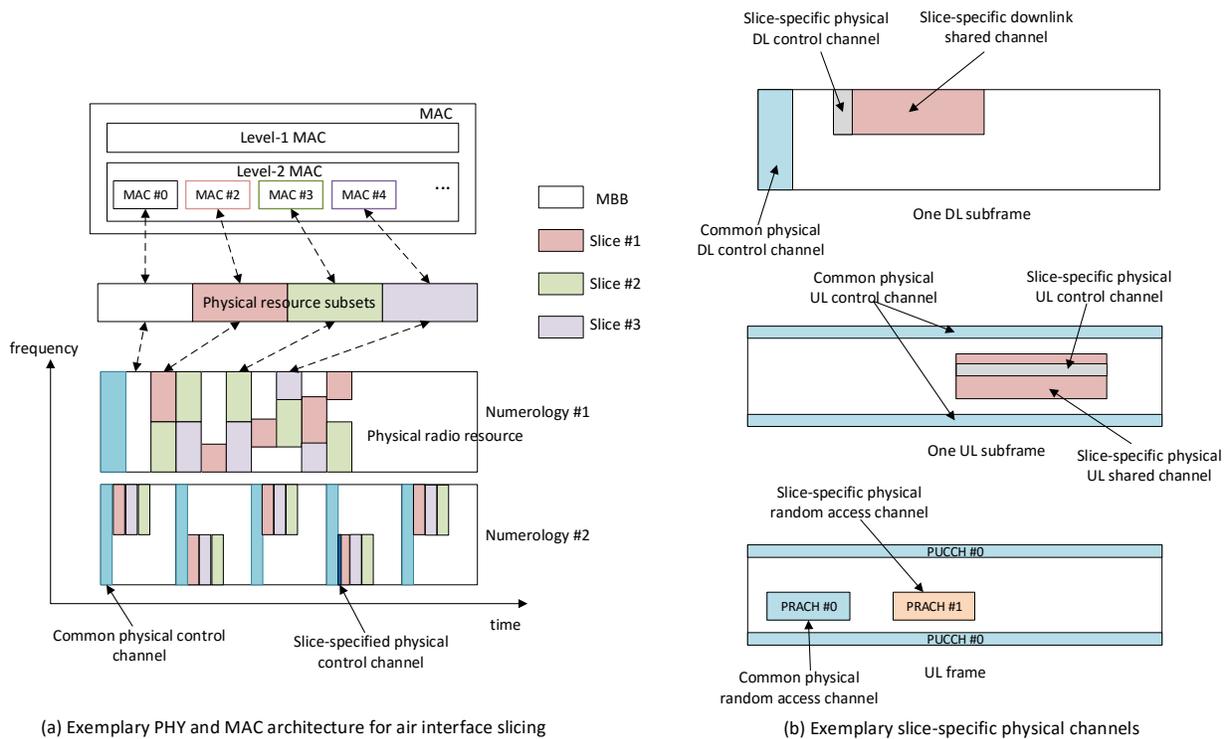

(a) Exemplary PHY and MAC architecture for air interface slicing

(b) Exemplary slice-specific physical channels

Figure 3 Example of air interface slicing

To identify a network slice, a network slice ID (sNetID) is assigned to the network slice. The sNetID is known by devices accessing the network slice and can be used to address all the devices in the network slice. The sNetID can be broadcast in the system information to indicate whether the slice is active in the BS.

For the key physical channels in the air interface, such as the physical downlink (DL) and uplink (UL) control channels, the physical random access channel and the physical shared channel, we can have slice-specific physical channels, as well as common physical channels. The common physical channels can be used by all slices, and the slice-specific physical channels are dedicated to the respective slices.

Figure 3 (b) shows an example of the physical downlink control channel in one DL subframe, the physical uplink control channel in one UL subframe, and the physical random access channel in one frame. The common physical DL control channel carries resource allocation information for the network slices. The sNetID is used to address the scheduled network slices. All the devices accessing a scheduled network slice can detect the common physical control information addressed to the corresponding sNetID. The slice-specific physical downlink control channel for a network slice is located in the radio resources assigned to the network slice. The slice-specific physical downlink control channel carriers scheduling information for the devices in the network slice. Similarly in the uplink, common and slice-specific physical uplink control channels can be designed. Devices accessing multiple network slices can aggregate the uplink control information and transmit it using the common physical control channel.

For random access, slice-specific random access channels can be used to differentiate the contention resolution and admission control of the network slices, so that, for instance, a crowded network slice with high random access collision probability should not affect devices accessing another network slice, or a network slice serving mission-critical services can get guaranteed low-latency access. The resource for the slice-specific random access channel of a slice can be indicated in downlink system broadcasting, which requires the slice to be active in the BS or the access point. For slices that have not been activated in the BS or access point, or for slices that do not require a slice-specific random access channel, the common random access channel can be used. In this case, the common random access channel can also be used as a way of activating a slice in the BS or access point.

### 4. Enabling network slicing in the RAN

As each slice can have its own RAN architecture, RAN operations such as mobile association, access control and load balancing schemes would be slice-specific instead of cell-specific as it is currently the case in mobile networks. Slice on/off operation would be enabled at each BS or access point. The control-plane (C-plane) and user-plane (U-plane) configuration could be tailored considering the slice-specific operation. In a sense, the slice-specific operation blurs the concept of physical cell site and makes the network operation more service/traffic/user oriented instead of physical cell oriented.

Depending on factors such as traffic type, traffic load and QoS requirement, the RAN architecture of each of the slices can be dynamically configured. For example, in one instance, slice #1 can only operate on a macro cell, slice #2 can only operate on small cells, and slice #3 can operate on both macro and small cells. In another situation, slice #1 could expand its operation to small cells, while slice #3 can terminate operation on some of the small cells. The slice-specific RAN architecture would require slice-specific control-plane/user-plane operation, slice on/off operation and slice-based treatment on access control and load balancing.

Three options on the control/user-plane configuration of the network slices can be considered: Option 1 refers to a case with common C-plane across network slices and dedicated U-plane for each of the slices.

Option 2 refers to a case with dedicated control and user planes for each of the slices. Option 3 refers to a case with common C-plane for all slices and dedicated C/U-plane for each of the slices. Some of the control-plane functions such as the functions in idle mode (e.g., paging, cell reselection, tracking area update) can be categorized into common C-plane slice functions, while the functions in connected mode (e.g., handover, dedicated bearer setup) can be categorized into slice-specific control plane functions. Option 2 allows slice-specific C-plane, but it may be costly to have always-on C-plane for each of the slice. Option 1 could require less design effort but lack the flexibility of tailoring the C-plane for each slice. Option 3 allows always-on common C-plane and tailored C-plane slice-specific C-plane.

The slice-specific RAN architecture inherently supports slice on/off, slice-based admission control and slice-specific load balancing. When turning on a slice, an access point would allocate radio resources for the slice and enable all radio and network functions associated with the slice, such as the and the corresponding physical channels. The triggers for turning on a slice at an access point could include: 1) Traffic load of that slice goes beyond a certain threshold; 2) The number of active devices operating on that slice goes beyond a certain threshold; 3) Need to keep service continuity; 4) Need to meet certain QoS requirements, such as low latency, ultra-reliability, etc. Slice-on at an access point can be triggered by a device or by the network. When triggered by a device, the network can decline the device slice-on request if, for example, the network assesses that the cost/overhead of serving the slice outweighs the service benefit.

Likewise, admission control and load balancing will be based on the availability of the requested slice at the BS or access point as well as the load conditions and the overall system performance. The system information of a BS or access point carries information on the active slices in the BS or access point. Based on the system information, a device can start the random access procedure with a BS that has the intended slice actively running. If the intended slice is not supported by a BS, the UE may still start the random access procedure considering factors such as link condition, QoS requirements, traffic load of the neighboring cells, etc. If the device makes the access request but the slice is not currently active in the BS or access point and the BS decided to accept the access request, the BS/access point will need to turn on the slice using the slice on/off procedure.

5. **Enabling network slicing in the CN**

Current core networks usually use dedicated hardware to support vertical networks. In 5G, the communication system is expected to support diverse services and applications with diverse requirements. The static, purpose-built vertical network as in current systems cannot address the use cases of the next decade and beyond. It calls for further simplification of the core network architecture and a more scalable design.

Network slicing allows the support of many industry and diverse use cases in one network by flexibly defining network functions, processing requirements, security procedures and execution nodes based on the requirements of each of the industry and use cases. NFV and SDN are technical enablers of network slicing in the CN. The SDN technology was mainly developed by the IT industry for efficient operation and management of datacenter and large IP and Ethernet networks. The goal of SDN is to separate the control plane from the data plane, and to make the control plane programmable through APIs in order to bring flexibility on how networks are deployed, operated and managed. NFV is primarily driven by network service providers. The goal of NFV is to virtualize network functions into software applications that can be run on industry off-the-shelf standard servers or as virtual machines running on those servers. NFV and SDN virtualize the network elements and functions to easily enable configured/reused network elements and functions in each slice to meet its own requirement. Figure 4 illustrates an example of specifically configured network slices built on common physical infrastructure comprised by various network elements. NFV virtualizes the physical network resources build on which SDN realizes the dynamic configuration of C-plane and U-plane of each network slices. Each slice could have tailored core network functions based on the targeted services and applications. To assign and configure the network functions for the network slices, a management plane could be needed.

Slicing in the core network, in the RAN and in the radio can follow different criteria. For instance, radio slicing can be based on communication requirements such latency, reliability, coverage, etc. Each of the radio slice takes a portion of the radio resource and defines one type of numerology. RAN slicing can be based on service/application as well as communication requirements. Each of the RAN slice could have tailored RAN architecture and C/U plane protocol. The CN slices can be defined to support different services/applications. Each of the CN slice could have tailored core network functions. Slice pairing functions are define to pair the radio, RAN, and CN slices to form end-to-end slices. The pairing among radio/RAN/CN slices can be 1:1 or 1:M, e.g., a radio slice could have multiple RAN slices built on top; a RAN slice could have multiple CN slices built on top.

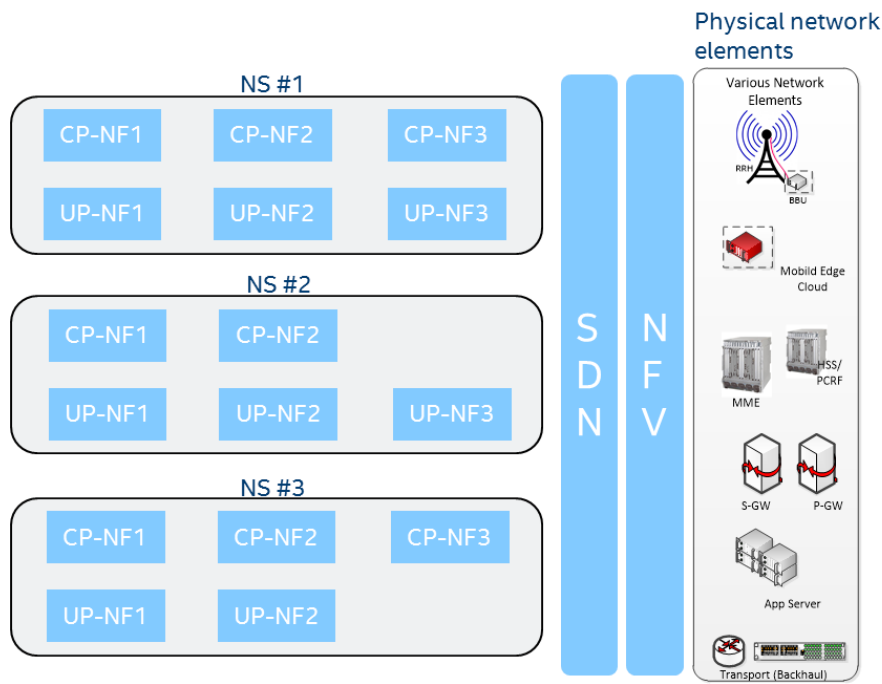

Figure 4 Flexible 5G framework using Network slices

## 6. Enhancing user experience with horizontal network slicing

Communication and computation have been helping each other in pushing the boundaries of information and computing technologies. At the network side, computation has been used to help communication by moving computation and storage to the edge. With edge cloud and edge computation, the communication link between the source and the destination is getting shorter, thereby improving the communication efficiency and reducing the amount of information propagation in the network. The optimal deployment of edge cloud and computation scheme varies. As a general rule, the less capable the end device is and/or the higher the device density, the more beneficial to put the cloud and computation closer to the network edge.

Moving forward at the device side, with the devices further shrinking in size from portable devices to wearable devices and the user expectation on computation keeping increasing, we expect communication will need to help computation to deliver the user experience, e.g., the network nodes slice out part of their computation resources to help computation at the portable device, while the portable devices slice out part of their computation resources to help the computation at the wearable devices. In this way, the network is horizontally sliced. The sliced out computation resources and the air interface connecting the two ends form an integrated part that delivers the required service. Figure 5 provides an example on illustrating the horizontal slicing concept. Computation resource slicing and the air interface supporting computation resource sharing need to be jointly designed for optimized performance. Figure 5 (a) illustrates the horizontal slicing concept.

Figure 5 (b) presents an example architecture on realizing computation slicing and resource sharing. The main building blocks are a communication module, a computation module, a management-plane (M-plane) module, the virtual machines (VM) formed by virtualization technique and the operation systems (OS) running on the VMs. Computation slicing is managed by the M-plane and implemented below OS. The M-planes monitors the system resource usage and the radio link condition. When the M-plane of the client sees benefit (in terms of performance, costs, etc.) of computation offloading or when the client OS requests resource beyond the client computation can support, it asks the host M-plane for sharing computation resource. The signaling exchange between the M-planes at the client and host are carried as air interface L3 signaling message. If the client's request for resource is accepted by the host, the client M-plane informs the client VM on the available of the resource. The client VM then slices the sliceable application based on the information from the M-plane and convey the generated executable code to the container engine, where container is developed to facilitate distribution and execution of sliceable application [8]. The container engine packs the sliced executable code into container and convey the data packet carrying the container to the communication module. The communication module packs the container data into L2 PDU and generate L1 data block and transmit to the host via the selected radio link. The communication module at the host decodes the received data blocks and convey to the container engine. The host container engine unpacks the received container and hands the executable code to the edge server VM. Upon completion of the computation task, the container engine of the host packs the executed results in the container and deliveries it back to the client container engine via communication link. The client container unpacks the executed result. The client OS then applies the received executed results. Figure 6 illustrates the described computation resource sharing procedure. Note that in the described example, the computation task slicing is implemented below OS. Alternatively, the slicing can be implemented at OS or above. In this case, the container information can be treated as normal user traffic and no special design is needed in the air interface. However, the performance (such as processing delay) will be affected by the OS performance. Implementing slicing at below OS is expected to give better performance. In this way, the host computing resource is virtualized into two parts. One part is used by its own applications, running on its own OS. Another part is allocated for client, and used by the client as remote resource. The application at the client has to be sliceable to be executed at two VMs simultaneously. One at the client, the other at the host. The client VM may serve as the master VM, and the host VM may serve as a co-processing engine to serve the client master VM. The traffic carrying the container can be made visible to the communication link. A dedicated L2 logical channel can be specified to carry the container traffic.

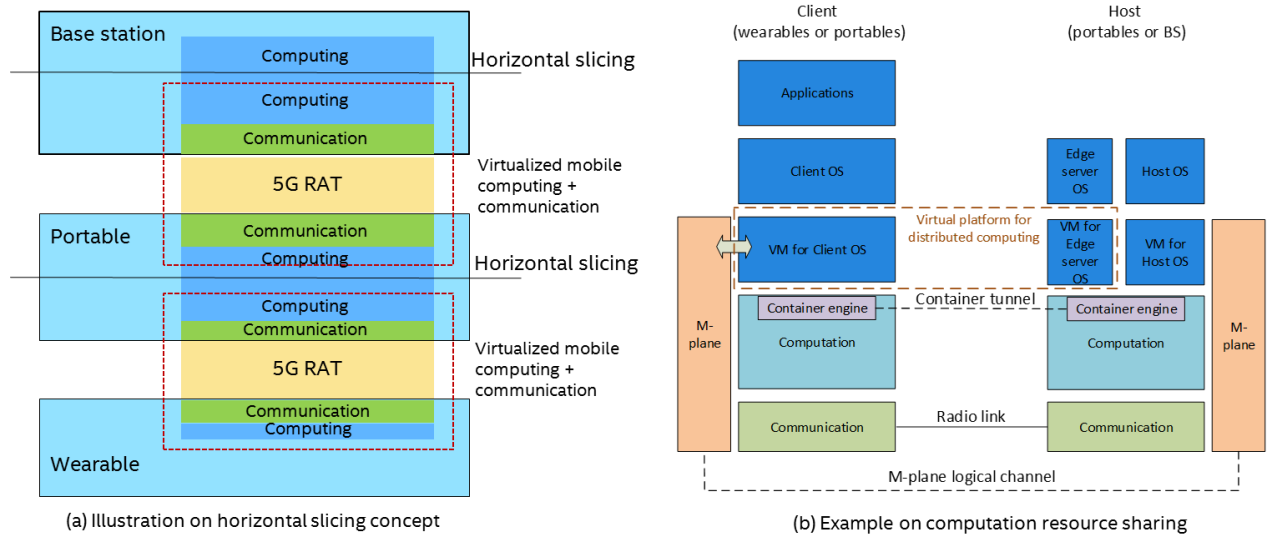

(a) Illustration on horizontal slicing concept

(b) Example on computation resource sharing

Figure 5 Illustration on enhancing device capability using horizontal slicing

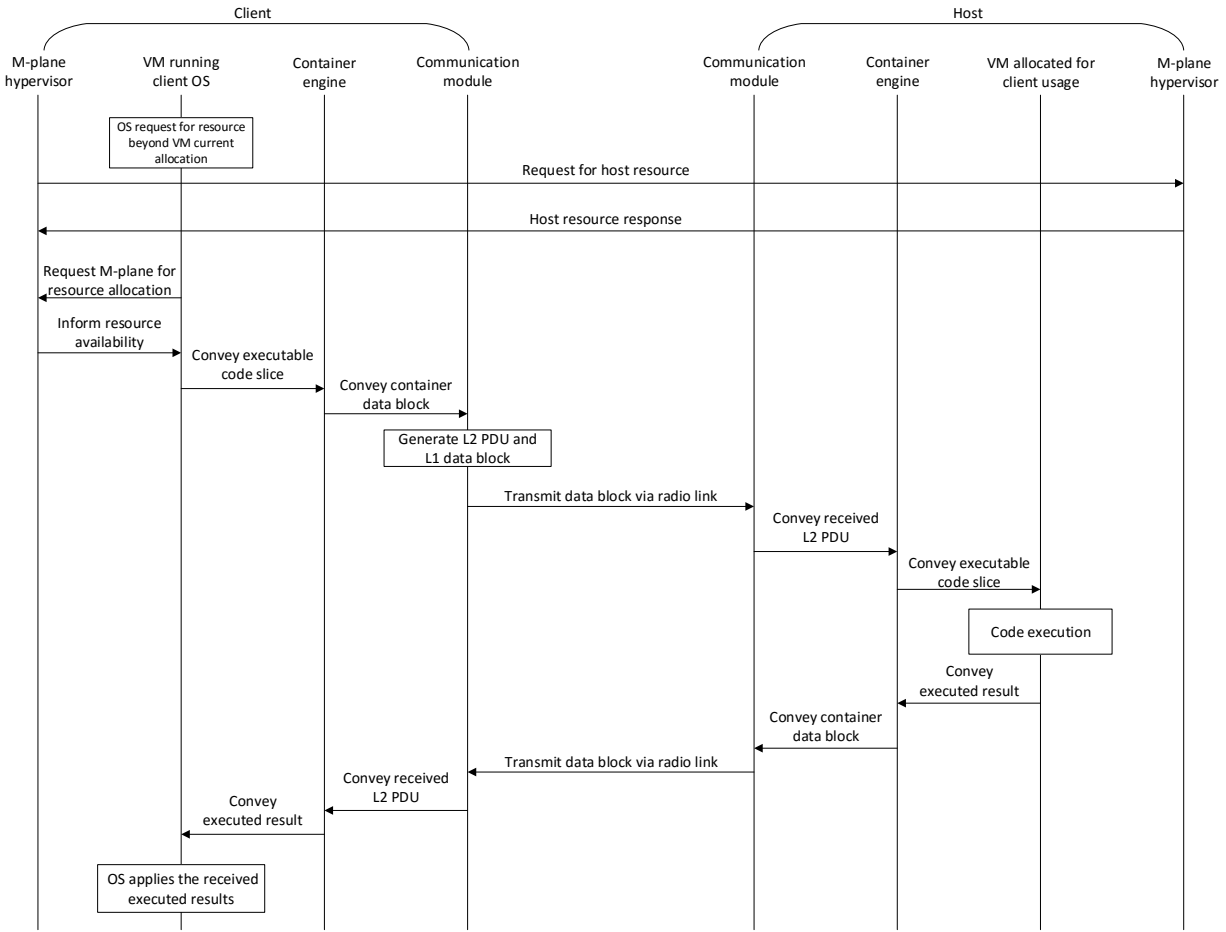

Figure 6 Computation resource sharing procedure

## 7. Global standardization on network slicing

The global standardization effort on network slicing is still in the early stage and is mostly focused on vertical slicing. Some on-going industry-wide studies on network slicing include the works in NGMN, 3GPP, 5GPPP Co-Funded framework, and WWRF. In 3GPP, network slicing is identified as one of the key technologies to be developed in 5G standardization and the study on core network slicing has been started [9]. The current discussions in the industry have been focused on the concept and requirements of network slicing, the ways and granularity of doing slicing, the impact of network slicing on the CN, the RAN and the devices, and the potential system architecture based on network slicing. We expect an industry-wide consensus on network slicing is to be developed down the road.

## 8. Conclusion and future research

In this paper, we presented a system framework for 5G wireless communication systems with vertical and horizontal network slicing. Vertical slicing enables flexible support of diverse services and applications.

Horizontal slicing augments mobile device capabilities and enhances user experiences. The 5G air interface is an integrated part of the network slicing framework and is enriched by network slicing. We discussed communication solutions in the CN, RAN and air interface and computation solutions on computation resource slicing and sharing under the proposed system framework. The computation and computation components in the system need to be jointly designed to achieve optimal communication-computation tradeoff.

As the industry moves towards vertical slicing as the phase-1 implementation, provision for enabling horizontal slicing as the phase-2 implementation is needed. Without a long-range vision, the industry may end up doing a version of Phase 1 network slicing and then realize the limitations when Phase 2 needs to be implemented. We aspire this paper serves the purpose of triggering the related discussions and technology development efforts.